\documentclass[onecolumn,nofootinbib]{revtex4}
\usepackage{graphicx} % Required for inserting images
\usepackage{setspace}
\usepackage{amsmath,amssymb}
\usepackage{graphicx}
\usepackage{dcolumn}
\usepackage{bm,color}
\usepackage{hyperref}
\usepackage{accents}
\usepackage{amssymb,float}
\usepackage{amsmath}
\usepackage{multirow}
\usepackage{tikz}
\usepackage[rightcaption]{sidecap}
\sidecaptionvpos{figure}{t}
\usepackage{enumerate}
\renewcommand{\arraystretch}{3}
\newcolumntype{P}[1]{>{\centering\arraybackslash}p{#1}}
\newcolumntype{M}[1]{>{\centering\arraybackslash}m{#1}}
\hypersetup{
    colorlinks=true,
    linkcolor=red,
    filecolor=magenta,      
    citecolor=blue
}
\newcommand{\udt}[3]{#1^{#2}_{\phantom{#2}#3}}

\newcommand{\dut}[3]{#1_{#2}^{\phantom{#2}#3}}

\newcommand{\lc}[1]{\accentset{\circ}{#1}}%Levi-Civita connection

\newcommand{\AN}[1]{{\bf\textcolor{blue}{(Abdurahmon)}}}

\begin{document}

\title{Cosmological Constraints on Minimal Cubic Galileon Models in Teleparallel Gravity }

\author{Akbar Davlataliev}
\email{akbar@astrin.uz}

\affiliation{School of Physics, Harbin Institute of Technology, Harbin 150001, People’s Republic of China}
\affiliation{University of Tashkent for Applied Sciences, Str. Gavhar 1, Tashkent 100149, Uzbekistan}
\affiliation{Department of Physics, New Uzbekistan University, Movarounnahr str. 1, Tashkent 100000, Uzbekistan}

\author{Abdurakhmon Nosirov}
\email{abdurahmonnosirov000203@gmail.com}
\affiliation{Center for Astronomy and Astrophysics, Center for Field Theory and Particle Physics and Department of Physics, Fudan University, 200438 Shanghai, China}

\author{Odil Yunusov}
\email{odilbekhamroev@gmail.com}
\affiliation{Center for Astronomy and Astrophysics, Center for Field Theory and Particle Physics and Department of Physics, Fudan University, 200438 Shanghai, China}
\date{\today}

\author{Bobomurat Ahmedov}%,\orcidlink{0000-0002-1232-610X}}
\email{ahmedov@astrin.uz}
\affiliation{School of Physics, Harbin Institute of Technology, Harbin 150001, People’s Republic of China}
\affiliation{Institute of Theoretical Physics, National University of Uzbekistan, Tashkent 100174, Uzbekistan}
\affiliation{Institute for Advanced Studies, New Uzbekistan University, Movarounnahr str. 1, Tashkent 100000, Uzbekistan}
\author{Jackson Levi Said}
\email{jackson.said@um.edu.mt}
\affiliation{Institute of Space Sciences and Astronomy, University of Malta, Malta, MSD 2080}
\affiliation{Department of Physics, University of Malta, Malta}

\begin{abstract}
Cubic Galileon cosmological models provide a well-motivated framework for investigating late-time cosmic acceleration beyond the standard $\Lambda$CDM paradigm. In this work, we study observational constraints on cubic Galileon models within the teleparallel gravity framework, where deviations from the standard teleparallel equivalent of general relativity are encoded through the model parameter $b_1$. We consider two scalar-field potentials, namely quadratic and exponential potentials, and analyze four representative scenarios: quadratic and exponential potentials with $b_1$ treated as a free parameter, together with the corresponding cases in which $b_1=2$ is fixed. Using the $\text{Pantheon}^+$ Type Ia supernova sample, cosmic chronometer measurements, SH0ES information, and baryon acoustic oscillation data, we constrain the cosmological and model parameters and compare the observational viability of the different scenarios. We find that the considered teleparallel cubic Galileon models can accommodate the late-time expansion history, although the statistical preference depends on the choice of potential and on whether $b_1$ is fixed or varied. In particular, the fixed-$b_1$ model with a quadratic potential provides the most competitive fit among the Galileon scenarios when BAO data are included, showing a lower $\chi^2_{\min}$ than $\Lambda$CDM and comparable support according to the AIC criterion. However, the BIC criterion continues to favor the minimal $\Lambda$CDM model because of the larger parameter space of the extended models. These results suggest that teleparallel cubic Galileon cosmologies remain phenomenologically viable, while a stronger claim regarding the Hubble tension requires further consistency tests.
\end{abstract}

\maketitle

\section{Introduction}
For decades, the $\Lambda$CDM framework has stood as the leading cosmological model, backed by overwhelming observational evidence across all scales ~\cite{Misner:1973prb,Clifton:2011jh}. In this picture, cold dark matter (CDM) provides gravitational stability for galaxies ~\cite{Baudis_2016,2005PhR...405..279B}, while the cosmological constant accounts for dark energy ~\cite{RevModPhys.75.559,Copeland:2006wr}. Nevertheless, despite considerable effort, the description of dark energy through a cosmological constant still faces internal coherence problems ~\cite{Weinberg:2000yb}, and no direct detection of dark matter particles has been achieved ~\cite{2004ARNPS..54..315G}. More recently, the validity of the $\Lambda$CDM model has been challenged by statistical discrepancies emerging from different cosmic surveys, commonly referred to as the \(H_0\) tension ~\cite{DiValentino:2020zio}. One interpretation of this mismatch contrasts late-time, model-independent measurements of the Hubble parameter ~\cite{2019ApJ...876...85R,H0LiCOW:2019pvv} with the early-universe predictions of the $\Lambda$CDM model ~\cite{Planck:2015fie,Planck:2018vyg}; alternatively, it could stem from systematic effects in certain observational methods ~\cite{Riess:2019qba,Pesce:2020xfe,2020MNRAS.496.3402D}. Ultimately, resolving the full scope of the tension may require entirely new approaches, such as gravitational wave standard sirens ~\cite{2019BAAS...51g..77T,2017arXiv170200786A,2019CQGra..36n3001B}.

The increasing observational challenges facing the $\Lambda$CDM paradigm ~\cite{Bernal:2016gxb,DiValentino:2020zio,DiValentino:2021izs} have motivated a renewed investigation into modifications of its core structure~\cite{Sotiriou:2008rp,Clifton:2011jh,CANTATA:2021asi}. Many such modifications introduce additional terms to the Einstein-Hilbert action, while still retaining the curvature-based description of gravity via the Levi-Civita connection ~\cite{Misner:1973prb,Nakahara:2003nw}. In parallel, a rising number of studies explore the possibility that torsion, rather than curvature, serves as the fundamental mechanism for gravitational interactions on differentiable manifolds~\cite{Bahamonde:2021gfp,Aldrovandi:2013wha,Cai:2015emx,Krssak:2018ywd}. Teleparallel gravity (TG) encompasses a broad class of theories where gravity arises from the torsion associated with a flat, metric-compatible connection—known as the teleparallel connection. This connection is curvature-free, meaning all curvature quantities vanish identically regardless of the metric components. A key consequence of adopting this connection is that the Ricci scalar computed from it is zero (\(R = 0\)), whereas the ordinary Ricci scalar \(\mathring{R}\) (where the circle denotes evaluation with the Levi-Civita connection) retains an arbitrary, non-zero value. In analogy with the Ricci scalar, TG introduces a torsion scalar \(T\). Up to a boundary term \(B\), this torsion scalar coincides with the ordinary Ricci scalar. Consequently, an action built linearly from the torsion scalar becomes dynamically indistinguishable from general relativity (GR)—a formulation known as the teleparallel equivalent of general relativity (TEGR).

The Cubic Galileon model \cite{Ye:2024,Nicolis:2008in,Deffayet:2009mn} belongs to the broader class of Galileon scalar-tensor theories, characterized by derivative self-interactions that preserve a Galilean shift symmetry in flat spacetime. Its cubic term, proportional to \(X\Box\phi\) where \(X = -\frac12 \partial_\mu\phi\partial^\mu\phi\), arises naturally in the Dvali–Gabadadze–Porrati (DGP) braneworld scenario \cite{Dvali:2000hr} and represents the simplest non-trivial Galileon interaction beyond the kinetic term. When embedded in a cosmological framework, the cubic Galileon can produce self-accelerating expansion without a cosmological constant, making it a viable dynamical dark energy candidate \cite{DeFelice:2011bh,Germani:2012ak}. Moreover, its kinetic braiding effect modifies the growth of structure and the propagation of gravitational potentials \cite{Kimura:2011km}, leaving distinctive imprints on both background and perturbation observables. Consequently, confronting Cubic Galileon models with the latest observational data—particularly supernovae, cosmic chronometers, baryonic acoustic oscillations, and local Hubble constant measurements—is essential to assess their viability and potential to resolve tensions such as the \(H_0\) discrepancy \cite{Frusciante:2019zkd,Zumalacarregui:2020bml,Brahma:2020feb,Ali:2021rgi}.

In addition to the public data sets, survey results can also be used in conjunction as priors to further analyze their consistency with said data sets. For instance, in Ref.~\cite{2019ApJ...876...85R} the SH0ES Team estimates the Hubble constant to be $73.30 \pm 1.04$ km s$^{-1}$ Mpc$^{-1}$ using Type Ia supernovae (SNIa), while the H0LiCOW Collaboration~\cite{H0LiCOW:2019pvv} reports $73.3^{+1.7}_{-1.8}$ km s$^{-1}$ Mpc$^{-1}$ based on strong lensing from quasars. One of the lowest local values comes from the tip of the red giant branch (TRGB) method, yielding $H_0 = 69.8 \pm 1.9$ km s$^{-1}$ Mpc$^{-1}$~\cite{Freedman:2019jwv}. Together with cosmic chronometers, SNIa, and baryonic acoustic oscillations, the impact of these priors on the most studied Cubic Galileon gravity models was recently examined in Ref.~\cite{Briffa:2021nxg} (where the analysis originally considered $f(T)$ gravity, but the methodology directly applies). The SNIa data set used in that study relied on the Pantheon release (PN), a compilation of 1048 SNIa relative luminosity distance measurements spanning $0.01 < z < 2.3$~\cite{Pan-STARRS1:2017jku}. More recently, the $\text{Pantheon}^+$ data set ($\text{PN}^+$\&SH0ES) has been released, building on Pantheon with 1701 events and a much higher concentration of low-redshift bins~\cite{Brout:2021mpj,Riess:2021jrx,Scolnic:2021amr}. This substantial increase in data points may yield significantly stronger constraints on cosmological models beyond $\Lambda$CDM, including Cubic Galileon gravity.

In this study, we carry out constraint analyses using the $\text{Pantheon}^+$\&SH0ES compilation ($\text{PN}^+$\&SH0ES) for the most viable Cubic Galileon gravity scenarios. The results are then contrasted with earlier analyses based on other observational data sets, allowing us to assess the differences in constraining power between the new $\text{PN}^+$\&SH0ES release and the previous Pantheon (PN) data set. We begin in Sec.~\ref{sec:2} with a brief review of key technical aspects of teleparallel gravity (TG) as it relates to the Cubic Galileon formulation. Sec.~\ref{sec:3} then describes the various data sets employed in our analysis. Our principal findings are presented in Sec.~\ref{sec:4}, where we derive constraints on the Cubic Galileon models using these data sets. In Sec.~\ref{sec:5}, we provide a comparison between our results and those of the standard $\Lambda$CDM cosmological model. Finally, Sec.~\ref{sec:6} summarizes our main conclusions and outlines possible directions for future research.

\section{Theoretical Setup}\label{sec:2}

\subsection{Teleparallel formulation of gravity}

Gravitational theories formulated in terms of spacetime curvature, such as General Relativity (GR), are based on the Levi--Civita connection, $\udt{\lc{\Gamma}}{\sigma}{\mu\nu}$ (where over-circles denote quantities constructed from it) being symmetric in lower indices. This connection defines the fundamental geometric objects of the theory, including the Riemann tensor, and enters directly into the Einstein--Hilbert action through the Ricci scalar. In contrast, Teleparallel Gravity (TG) provides an equivalent description in which curvature is replaced by torsion, and the Levi--Civita connection is substituted by the teleparallel connection $\udt{\Gamma}{\sigma}{\mu\nu}$ \cite{Aldrovandi:2013wha,Bahamonde:2021gfp,Cai:2015emx,Krssak:2018ywd}.

The mathematical structures underlying these two formulations differ substantially. While curvature-based theories employ the metric tensor $g_{\mu\nu}$ as the fundamental variable, TG is formulated in terms of the tetrad field $\udt{e}{A}{\mu}$ together with a flat spin connection $\udt{\omega}{B}{C\nu}$. Greek indices refer to spacetime coordinates, whereas Latin indices label components in the local Minkowski frame. Although tetrads and spin connections can also be introduced in GR, their role there is less direct. In TG, by contrast, the spin connection encodes inertial effects only. The tetrad is related to the metric via
\begin{align}
    g_{\mu\nu} = \udt{e}{A}{\mu}\udt{e}{B}{\nu}\eta_{AB}, 
    \qquad 
    \eta_{AB} = \dut{E}{A}{\mu}\dut{E}{B}{\nu} g_{\mu\nu},
    \label{eq:metr_trans}
\end{align}
where $\dut{E}{A}{\mu}$ denotes the inverse tetrad. This relation reflects the non-uniqueness of the tetrad, with the spin connection ensuring invariance under local Lorentz transformations.

The tetrad--spin connection pair fully determines the teleparallel connection, which can be written as \cite{Cai:2015emx,Krssak:2018ywd}
\begin{equation}
    \Gamma^{\lambda}{}_{\nu\mu}
    =
    \dut{E}{A}{\lambda}\partial_{\mu}\udt{e}{A}{\nu}
    +
    \dut{E}{A}{\lambda}\udt{\omega}{A}{B\mu}\udt{e}{B}{\nu}.
\end{equation}
The flatness of the spin connection is imposed through
\begin{equation}
    \partial_{[\mu}\udt{\omega}{A}{|B|\nu]} 
    + 
    \udt{\omega}{A}{C[\mu}\udt{\omega}{C}{|B|\nu]} 
    \equiv 0,
\end{equation}
which guarantees vanishing curvature. For suitable choices of tetrads, one can adopt the Weitzenb\"{o}ck gauge \cite{Weitzenbock:1923efa}, in which the spin connection vanishes identically.

In TG, gravitational dynamics are encoded in torsion rather than curvature. The Riemann tensor constructed from the teleparallel connection vanishes identically,
\begin{equation}
    \udt{R}{\alpha}{\beta\gamma\epsilon}(\udt{\Gamma}{\sigma}{\mu\nu}) \equiv 0,
\end{equation}
while the Levi--Civita Riemann tensor remains non-zero. The fundamental field strength is given by the torsion tensor \cite{Aldrovandi:2013wha,Ortin:2004ms}
\begin{equation}
    \udt{T}{A}{\mu\nu} := 2\udt{\Gamma}{A}{[\nu\mu]},
\end{equation}
which is invariant under both diffeomorphisms and local Lorentz transformations \cite{Bahamonde:2021gfp,Krssak:2015oua}.

The torsion tensor can be decomposed into three irreducible components: an axial vector $a_{\mu}$, a vector $v_{\mu}$, and a purely tensorial part $t_{\lambda\mu\nu}$ \cite{Hayashi:1979qx,Bahamonde:2017wwk,Bahamonde:2024zkb}. From these, one constructs three independent scalar invariants,
\begin{align}
    T_{\text{ax}} &:= a_{\mu}a^{\mu}, \\
    T_{\text{vec}} &:= v_{\mu}v^{\mu}, \\
    T_{\text{ten}} &:= t_{\lambda\mu\nu}t^{\lambda\mu\nu},
\end{align}
which span all parity-even quadratic contractions of the torsion tensor \cite{Bahamonde:2015zma}.

A particular linear combination of these invariants defines the torsion scalar $T$, which plays a central role in TG. Importantly, it is related to the Ricci scalar of GR up to a total divergence term. Since the Ricci scalar computed from the teleparallel connection vanishes, one finds
\begin{equation}
    \lc{R} = -T + B,
\end{equation}
where $B$ is a boundary term and $e = \det(\udt{e}{A}{\mu}) = \sqrt{-g}$. This identity establishes the equivalence between TG and GR at the level of the action.

The linear torsion scalar action corresponds to the Teleparallel Equivalent of General Relativity (TEGR), which is dynamically equivalent to GR \cite{Hehl:1994ue,Aldrovandi:2013wha}. Extending this framework, one can generalize the Lagrangian to an arbitrary function $f(T)$ \cite{Ferraro:2006jd,Ferraro:2008ey,Bengochea:2008gz,Paliathanasis:2017htk,Linder:2010py,Chen:2010va,Bahamonde:2019zea,Cai:2015emx,Farrugia:2016qqe,Iorio:2012cm,Deng:2018ncg}. A notable advantage of $f(T)$ gravity over its curvature-based counterpart $f(R)$ is that the resulting field equations remain second order.

\subsection{Cubic Galileon extension in teleparallel gravity}

In this work, we consider a Galileon scalar field coupled to gravity within the teleparallel framework. Matter fields are minimally coupled in the same manner as in GR, by promoting partial derivatives to covariant derivatives constructed from the Levi--Civita connection,
\begin{equation}
    \partial_{\mu} \rightarrow \mathring{\nabla}_{\mu},
\end{equation}
which applies exclusively to the matter sector.

We adopt the following action for cubic Galileon gravity \cite{Ye:2024}:
\begin{align}
    S = \int e \left[ \frac{1}{2\kappa c} f(T) + X  - \xi X \Box \phi - \frac{3 H_0^2}{c^2 \kappa}V(\phi) \right] \mathrm{d}^4 x + S_m \, ,
    \label{action}
\end{align}
where
\begin{align}
    f(T) &= -T + \alpha (-T)^{b_1} \, , \\
    X &= -\frac{1}{2} \partial^\mu \phi \, \partial_\mu \phi \, ,
\end{align}
and \(V(\phi)\) denotes the scalar-field potential.

The matter action is given by
\begin{align}
    S_m = \frac{1}{c} \int e \, \mathcal{L}_m \, \mathrm{d}^4 x \, .
\end{align}
Its variation with respect to the tetrad yields
\begin{align}
    \delta S_m 
    &= \frac{1}{2c} \int e \, T^{\mu\nu} \, \delta g_{\mu\nu} \, \mathrm{d}^4 x \nonumber \\
    &= \frac{1}{c} \int e \, \eta_{ab} \, T^{\mu\nu} \, e^{a}{}_{\mu} \, \delta e^{b}{}_{\nu} \, \mathrm{d}^4 x \, .
\end{align}

Throughout this work, we adopt the metric signature $(+,-,-,-)$. The energy--momentum tensor for a perfect fluid is
\begin{align}
    T^{\mu\nu} = (\rho + P)\frac{u^\mu u^\nu}{c^2} - P g^{\mu\nu} \, .
\end{align}

We consider the spatially flat FLRW tetrad
\begin{align}
    e^{a}{}_{\mu} = \mathrm{diag}\left(1, a(t), a(t), a(t)\right) \, ,
\end{align}
which reproduces the metric via
\begin{align}
    g_{\mu\nu} = \eta_{ab} \, e^{a}{}_{\mu} \, e^{b}{}_{\nu} \, .
\end{align}

\subsection{Field equations}

Varying the action \eqref{action} with respect to the tetrad leads to the modified Friedmann equations. The first Friedmann equation reads
\begin{align}
    H^2
    = -\frac{1}{6} \mathcal{A}(H)
    + \frac{\kappa}{3}\rho
    - \frac{c\kappa}{3}V(\phi)
    + \frac{c\kappa}{6}\dot{\phi}^{\,2}
    - c\kappa\,\xi\,H\,\dot{\phi}^{\,3}\,,
    \label{friedmann1}
\end{align}
where we have defined
\begin{align}
    \mathcal{A}(H) \equiv \alpha \,6^{\,b_1}(2b_1-1)\,H^{2b_1}\,.
\end{align}

The second Friedmann equation takes the form
\begin{align}
    \left(12H^2 + 2b\,\mathcal{A}\right)\dot{H}
    + H^2\Big[
        18H^2 + 3\mathcal{A}
        + 2\kappa P
        + 3c\kappa\big(2V(\phi) + \dot{\phi}^{\,2}(1+2\xi\ddot{\phi})\big)
    \Big]
    = 0 \, .
    \label{friedmann2}
\end{align}
Solving explicitly for $\dot{H}$, one obtains
\begin{align}
    \dot{H}
    = -\frac{H^2\Big[
        18H^2 + 3\mathcal{A}
        + 2\kappa P
        + 3c\kappa\big(2V(\phi) + \dot{\phi}^{\,2}(1+2\xi\ddot{\phi})\big)
    \Big]}{12H^2 + 2b_1\,\mathcal{A}} \, .
    \label{friedmann2_solved}
\end{align}

Variation of the action with respect to the scalar field $\phi$ yields the equation of motion
\begin{align}
    9 \xi H^2 \dot{\phi}^2
    + 3 \xi \dot{H} \dot{\phi}^2
    + V'(\phi)
    - \ddot{\phi}
    + 3 H \dot{\phi} \left( -1 + 2 \xi \ddot{\phi} \right)
    = 0 \, .
    \label{scalar_eom}
\end{align}
Here, overdots and primes denote derivatives with respect to cosmic time and the scalar field, respectively,
\begin{equation}
    \dot{H} \equiv \frac{dH}{dt}, 
    \quad 
    \dot{\phi} \equiv \frac{d\phi}{dt}, 
    \quad 
    \ddot{\phi} \equiv \frac{d^2\phi}{dt^2}, 
    \quad 
    V'(\phi) \equiv \frac{dV}{d\phi}.
\end{equation}

\section{Observational Data Analysis}\label{sec:3}

\subsection{Cosmic Chronometers (CC) Measurements}

To constrain the Hubble expansion rate, we employ a compilation of thirty-one Cosmic Chronometer (CC) measurements~\cite{Zhang:2012mp,Jimenez:2003iv,2016JCAP...05..014M,Simon:2004tf,Moresco:2012jh,Stern:2009ep,2015MNRAS.450L..16M}. The CC approach relies on spectroscopic age determinations of massive, passively evolving galaxies. This technique allows direct estimation of the Hubble parameter at different redshifts, extending to approximately $z \lesssim 2$. A key advantage of this method is that it does not depend on any assumed cosmological background model nor on the Cepheid distance ladder. Nevertheless, it requires stellar population synthesis modeling in order to determine galaxy ages reliably.

The method is based on measuring the differential age evolution between two galaxies located at nearby redshifts. From the observable quantity $\Delta z / \Delta t$, one can directly evaluate the Hubble parameter through

\begin{equation}
H(z) = -\frac{1}{1+z}\frac{\Delta z}{\Delta t}.
\end{equation}

Compared to techniques that rely on absolute age estimates, the differential nature of the CC method significantly reduces systematic uncertainties~\cite{Jimenez:2001gg}.

The chi-square function associated with the CC dataset is defined as

\begin{equation}
\chi^2_H = \sum_{i=1}^{31}
\frac{\left[ H(z_i,\Theta) - H_{\mathrm{obs}}(z_i) \right]^2}
{\sigma_H^2(z_i)},
\label{eq:chi_cc}
\end{equation}

where $H(z_i,\Theta)$ represents the theoretical prediction for a given parameter set $\Theta$, $H_{\mathrm{obs}}(z_i)$ denotes the measured value, and $\sigma_H(z_i)$ is the corresponding uncertainty.

\subsection{Type Ia Supernovae Observations}

In addition to CC data, we include Type Ia Supernovae (SNIa) observations in our Markov Chain Monte Carlo analysis to be aplied. These events originate from thermonuclear explosions in binary systems and are characterized by a nearly uniform intrinsic luminosity, which makes them reliable standard candles for cosmological distance measurements.

The observable quantity is the distance modulus, defined as the difference between the apparent magnitude $m$ and the absolute magnitude $M$. For a source at redshift $z_i$, it is given by

\begin{equation}
\mu(z_i,\Theta) = m - M = 5 \log_{10}\left[D_L(z_i,\Theta)\right] + 25,
\end{equation}

where the luminosity distance is expressed as

\begin{equation}
D_L(z,\Theta) = c(1+z) \int_0^z \frac{dz'}{H(z',\Theta)}.
\end{equation}

Since the absolute magnitude $M$ is not independently determined, it is treated as a nuisance parameter and marginalized over during the statistical analysis. Model predictions for $\mu(z)$ are compared with observational data, and parameter constraints are obtained by minimizing the likelihood function

\begin{equation}
\chi^2_{\mathrm{SN}} =
\Delta \mu^T C^{-1} \Delta \mu,
\label{eq:chi_sn}
\end{equation}

where $\Delta \mu = \mu(z_i,\Theta) - \mu_{\mathrm{obs}}(z_i)$ and $C$ denotes the covariance matrix that incorporates both statistical and systematic uncertainties \cite{SNLS:2011lii}.

In this work, we utilize $\text{Pantheon}^+$ ($\text{PN}^+$\&SH0ES) \cite{Scolnic:2021amr} SNIa compilation, extended from the Pantheon sample (PN) \cite{Pan-STARRS1:2017jku}. The original Pantheon dataset consists of 1048 SNIa events, whereas $\text{Pantheon}^+$ expands this number to 1701 supernovae, thereby improving statistical precision. The $\text{Pantheon}^+$\&SH0ES analysis further incorporates Cepheid-based distance calibrations from the SH0ES program (R22 \cite{2022ApJ...934L...7R}), which helps break the degeneracy between the absolute magnitude $M$ and the Hubble constant $H_0$. Moreover, $\text{Pantheon}^+$ spans a broader redshift interval, $0.01 < z < 2.5$, allowing for a more comprehensive control of systematic effects compared to the original Pantheon compilation.

\subsection{Baryon Acoustic Oscillations Measurements}

We also include Baryon Acoustic Oscillation (BAO) measurements from multiple independent surveys. The dataset combines results from the SDSS Main Galaxy Sample at $z_{\mathrm{eff}} = 0.15$ \cite{Ross:2014qpa}, the 6dF Galaxy Survey at $z_{\mathrm{eff}} = 0.106$ \cite{2011MNRAS.416.3017B}, and the BOSS DR11 quasar Lyman-$\alpha$ sample at $z_{\mathrm{eff}} = 2.4$ \cite{BOSS:2017uab}. Additionally, we incorporate angular diameter distance and Hubble rate measurements from SDSS-IV eBOSS DR14 quasars at $z_{\mathrm{eff}} = \{0.98, 1.23, 1.52, 1.94\}$ \cite{eBOSS:2018yfg}, together with the SDSS-III BOSS DR12 consensus measurements at $z_{\mathrm{eff}} = \{0.38, 0.51, 0.61\}$ \cite{2017MNRAS.470.2617A}. For the latter datasets, the full covariance matrices are taken into account.

The relevant cosmological distance measures are

\begin{equation}
D_H(z) = \frac{c}{H(z)}, \quad
D_M(z) = (1+z) D_A(z), \quad
D_V(z) = \left[(1+z)^2 D_A^2(z) \frac{cz}{H(z)}\right]^{1/3},
\end{equation}
where the angular diameter distance satisfies

\begin{equation}
D_A(z) = \frac{D_L(z)}{(1+z)^2}.
\end{equation}

From the observational BAO data, we construct the combinations involving the comoving sound horizon at the drag epoch $r_s(z_d)$, evaluated at $z_d \approx 1059.94$ \cite{2020A&A...641A...6P}. 

The sound horizon is computed via

\begin{equation}
r_s(z) = \int_z^\infty \frac{c_s(\tilde{z})}{H(\tilde{z})} d\tilde{z}
= \frac{1}{\sqrt{3}} \int_0^{1/(1+z)}
\frac{da}{a^2 H(a) \sqrt{1 + \left(\frac{3\Omega_{b,0}}{4\Omega_{\gamma,0}}\right)a}},
\end{equation}
where we adopt $\Omega_{b,0} = 0.02242$ \cite{2020A&A...641A...6P}, $T_0 = 2.7255\,\mathrm{K}$ \cite{2009ApJ...707..916F}, and a fiducial value $r_{s,\mathrm{fid}}(z_d) = 147.78\,\mathrm{Mpc}$.

The chi-square contribution from BAO measurements is written as

\begin{equation}
\chi^2_{\mathrm{BAO}}(\Theta)
=
\Delta G^T C_{\mathrm{BAO}}^{-1} \Delta G,
\label{eq:chi_bao}
\end{equation}
where $\Delta G(z_i,\Theta) = G(z_i,\Theta) - G_{\mathrm{obs}}(z_i)$ and $C_{\mathrm{BAO}}$ is the total covariance matrix associated with the selected BAO observations.

\section{Results and Discussions}\label{sec:4}

In this section, we present and discuss the outcomes of our analysis, following the methodology described in Sec.~\ref{sec:3} and employing the observational data sets introduced earlier. Each subsection concentrates on the most competitive Cubic Galileon scenarios, displaying contour plots of the fitted parameters with $1\sigma$ and $2\sigma$ confidence levels, accompanied by tables summarizing the final numerical results. These models have attracted considerable attention in the literature and are frequently investigated because they successfully reproduce key features of our cosmological history. In all tables and posterior plots, we report the Hubble constant $H_0$, the present-day matter density parameter $\Omega_{m,0}$, along with the relevant model parameters. This allows us to examine how different independent data sets and cosmological models influence the $H_0$ tension.

We consider the following four distinct models, which differ in the treatment of the parameter $b_1$ and the choice of the potential $V(\phi)$:

\begin{itemize}
    \item \textbf{Model I:} Fixed $b_1 = 2$ with a power-law-like potential $V(\phi) = c_2 \phi^2$.
    \item \textbf{Model II:} Fixed $b_1 = 2$ with an exponential potential $V(\phi) = c_2 e^{c_1 \phi}$.
    \item \textbf{Model III:} Free $b_1$ (to be constrained by data) with a power-law-like potential $V(\phi) = c_2 \phi^2$.
    \item \textbf{Model IV:} Free $b_1$ (to be constrained by data) with an exponential potential $V(\phi) = c_2 e^{c_1 \phi}$.
\end{itemize}
The power-law potential $V \propto \phi^2$ is chosen as the simplest  renormalizable mass term, often used as a benchmark for dark energy tracker solutions~\cite{Zlatev:1998tr}, and it has been widely adopted in Galileon cosmology to isolate the effects of kinetic self-interactions~\cite{DeFelice:2010nf}. The exponential potential $V \propto e^{c_1\phi}$ is motivated by its natural appearance in high-energy physics and its ability to produce scaling solutions that ease the coincidence problem~\cite{Copeland:1997et, Ferreira:1997hj}; it also serves as a standard test case for cubic Galileon models~\cite{Nesseris:2010pc}.

In the subsequent subsections, we present the detailed results obtained for each case.

\subsection{Model I}
As a reference scenario (Model I), we fix $b_1 = 2$, corresponding to a quadratic correction in the torsion scalar. This choice allows us to isolate the impact of a non-linear $f(T)$ contribution and to provide a benchmark for comparison with the fully free $b_1$ case.

\begin{figure}
    \centering
    \includegraphics[width=0.8\linewidth]{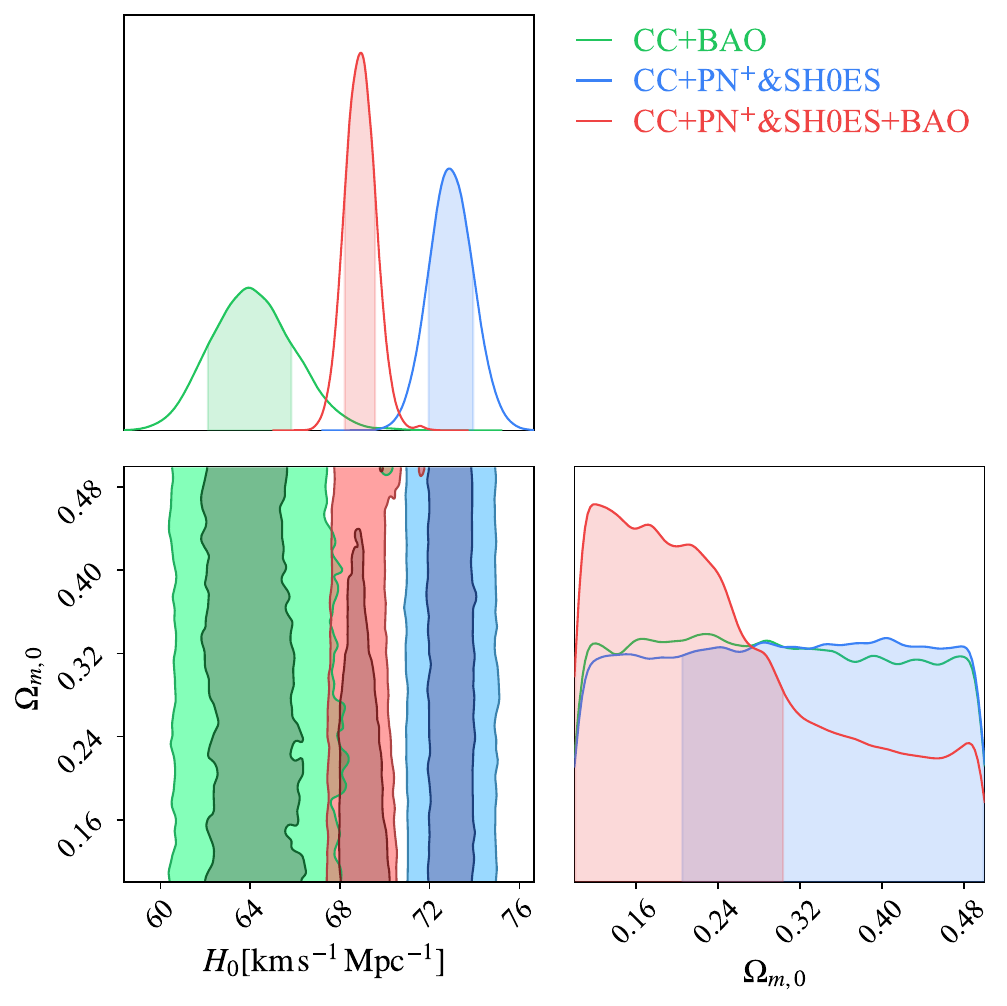}
    \caption{MCMC corner plot for the cubic Galileon teleparallel cosmology with quadratic effective potential $V(\phi)=c_2 \phi^2$ and the parameter fixed to $b_1=2$. The contours show the 68\% and 95\% confidence levels of the two-dimensional marginalized posterior distributions, and the diagonal panels display the corresponding one-dimensional marginalized posteriors.}
    \label{fig:model1}
\end{figure}

The constraints obtained for the parameters of Model I are displayed in Fig.~\ref{fig:model1}. This figure presents both the confidence contours and the posterior distributions for various combinations of observational data sets, specifically comparing results using only the cosmic chronometer (CC) catalog versus those using CC combined with $\text{Pantheon}^+$\&SH0ES (CC+$\text{PN}^+$\&SH0ES). A closer inspection of the posteriors reveals that the parameter uncertainties become considerably tighter when $\text{PN}^+$\&SH0ES data are included, with the Hubble constant $H_0$ showing a particularly marked improvement in precision. When baryonic acoustic oscillation (BAO) data are added, the previously present degeneracy is broken, revealing a clear anti‑correlation between the parameters. It is worth noting that the CC+$\text{PN}^+$\&SH0ES combination exhibits a degeneracy between $\Omega_{m,0}$ and $H_0$, whereas an anti‑correlation between $b_1$ and $\Omega_{m,0}$ appears for all data set combinations. However, the strength of this anti‑correlation weakens when BAO data are included.

The best‑fit values for the cosmological parameters, the model parameter $b_1$, and the nuisance parameter $M$ are summarized in Table~\ref{tab11}. The results clearly show that $H_0$ estimates obtained from data combinations that include $\text{PN}^+$\&SH0ES are systematically higher than those derived from CC alone. This trend is consistent with the locally measured value reported by the SH0ES team (R22), namely $H_0 = 73.30 \pm 1.04$ km s$^{-1}$ Mpc$^{-1}$~\cite{Riess:2021jrx}. Among all combinations, the highest $H_0$ value is found for CC+$\text{PN}^+$\&SH0ES, yielding $H_0 = 72.89^{+1.06}_{-0.93}$ km s$^{-1}$ Mpc$^{-1}$.

\begin{table}[htbp]\small
    \centering
    \begin{tabular}{l c c c c c c c c}
        \hline\hline
        Data sets 
        & $H_0 \,[\mathrm{km\, s^{-1}\, Mpc^{-1}}]$ & $\Omega_{m,0}$ & $\phi_0$ & $\dot{\phi}_0$ & $c_2$ & $M$ \\
        \hline
        ${\rm CC}$&$68.93^{+0.57}_{-0.75}$&$0.13^{+0.247}_{-0.029}$&$-$&$-$&$-4.09^{+1.34}_{-0.81}$&$-$\\
        ${\rm CC+BAO}$&$63.9^{+1.9}_{-1.8}$&$-$&$-$&$-$&$-4.37^{+1.98}_{-0.61}$&$-$\\
         ${\rm CC+PN^+\&SH0ES}$&$72.89^{+1.06}_{-0.93}$&$0.405^{+0.094}_{-0.200}$&$-1.99^{+0.92}_{-0.00}$&$-$&$-4.47^{+2.18}_{-0.52}$&$-19.258^{+0.029}_{-0.029}$\\
         ${\rm CC+PN^+\&SH0ES+BAO}$&$68.92^{+0.65}_{-0.70}$&$0.10^{+0.20}_{-0.00}$&$-$&$0.040^{+0.243}_{-0.044}$&$-$&$-19.368^{+0.020}_{-0.015}$\\
        \hline\hline
    \end{tabular}
    \caption{Best-fit values of the cosmological parameters obtained from different observational data sets for the potential $V(\phi)=c_2 \phi^2$ ($b_1=2$).}
    \label{tab11}
\end{table}

\subsection{Model II}
\begin{figure}
    \centering
    \includegraphics[width=0.8\linewidth]{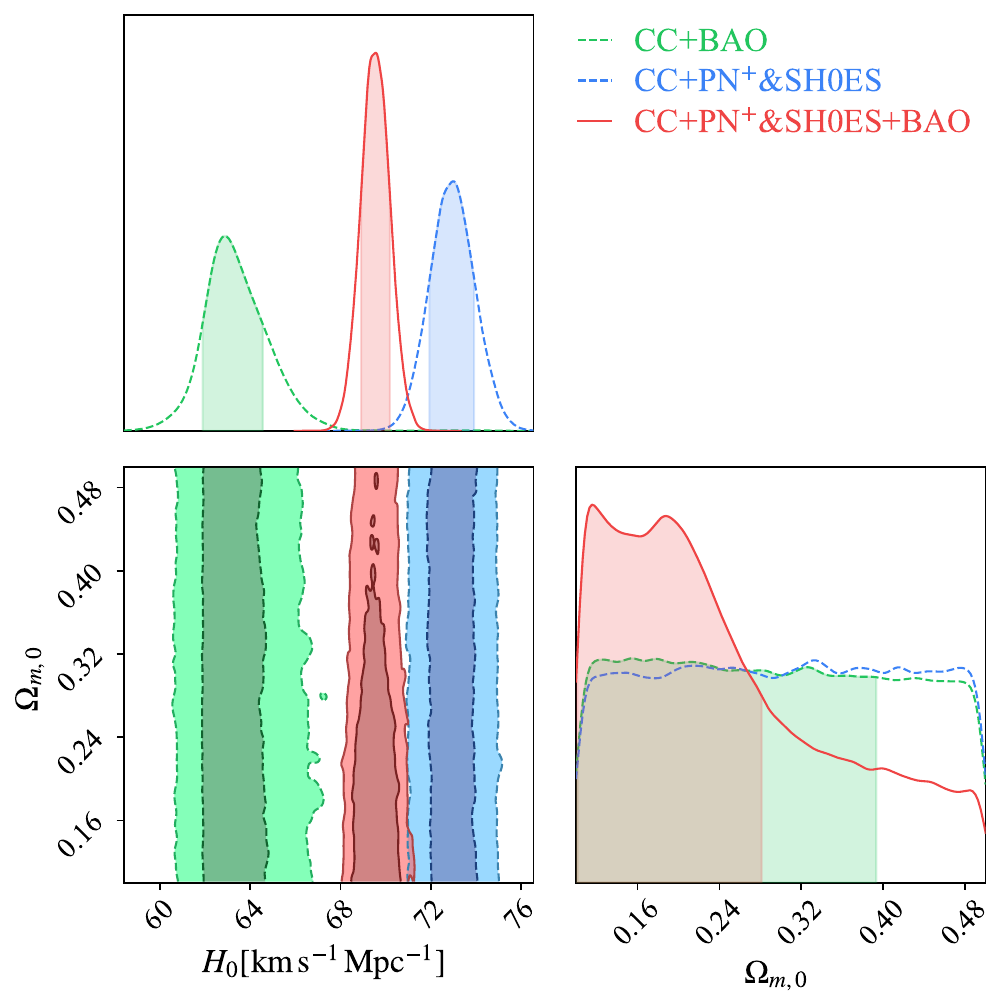}
    \caption{MCMC corner plot for the cubic Galileon teleparallel cosmology with exponential effective potential $V(\phi) = c_2 e^{c_1 \phi}$ and $b_1=2$ fixed. The two-dimensional contours correspond to the 68\% and 95\% confidence levels of the marginalized posterior distributions, while the diagonal panels give the one-dimensional marginalized posteriors.}
    \label{fig:model2}
\end{figure}

Model II, defined by the exponential effective potential \(V(\phi)=c_2 e^{c_1\phi}\) with \(b_1=2\) fixed, was constrained using four combinations of data. The key difference with respect to Model I lies in the choice of scalar-field potential: here we consider an exponential potential, enabling us to investigate how the cosmological dynamics depend on the form of the potential within the same modified gravity setup. The corner plots for Model II are displayed in Fig.~\ref{fig:model2}. The best-fit values in Table~\ref{tab12} show that the Hubble constant \(H_0\) is sensitive to the inclusion of SH0ES data: the CC+$\text{PN}^+$\&SH0ES and the full CC+$\text{PN}^+$\&SH0ES+BAO combinations push \(H_0\) above \(69\,\mathrm{km\,s^{-1}Mpc^{-1}}\), partially easing the Hubble tension, whereas CC alone and CC+BAO prefer lower values. The matter density parameter \(\Omega_{m,0}\) is poorly constrained except when BAO data are included; the scalar-field initial conditions \(\phi_0\) and \(\dot{\phi}_0\) remain consistent with zero within \(1\sigma\) across all datasets, indicating no strong dynamical evolution is required. The coupling parameter \(c_1\) is essentially unconstrained in all cases, while \(c_2\) only becomes well-measured when BAO data are added. The absolute magnitude \(M\) is tightly constrained to around \(-19.3\) to \(-19.4\) in the SNIa combinations.

\begin{table}[htbp]\small
    \centering
    \begin{tabular}{l c c c c c c c c}
        \hline\hline
        Data sets 
        & $H_0 \,[\mathrm{km\, s^{-1}\, Mpc^{-1}}]$ & $\Omega_{m,0}$ & $\phi_0$ & $\dot{\phi}_0$ & $c_1$ & $c_2$ & $M$ \\
        \hline
         ${\rm CC}$&$63.2^{+5.3}_{-5.0}$&$0.482^{+0.017}_{-0.273}$&$0.1^{+1.2}_{-1.5}$&$0.0^{+1.3}_{-1.2}$&$-0.5^{+6.8}_{-6.7}$&$-$&$-$\\
         ${\rm CC+BAO}$&$62.83^{+1.74}_{-0.95}$&$-$&$0.0^{+1.2}_{-1.4}$&$-$&$-$&$-9.08^{+7.90}_{-0.86}$&$-$\\
         ${\rm CC+PN^+\&SH0ES}$&$73.04^{+0.89}_{-1.10}$&$-$&$-$&$-0.09^{+0.45}_{-0.27}$&$1.2^{+2.9}_{-5.2}$&$-$&$-19.255^{+0.027}_{-0.031}$\\
         ${\rm CC+PN^+\&SH0ES+BAO}$&$69.57^{+0.62}_{-0.66}$&$0.1^{+0.18}_{-0.00}$&$-0.53^{+1.65}_{-0.55}$&$-0.00^{+0.29}_{-0.31}$&$-$&$-$&$-19.367^{+0.021}_{-0.014}$\\
        \hline\hline
    \end{tabular}
    \caption{Best-fit values of the cosmological parameters obtained from different observational data sets for the potential $V(\phi) = c_2 e^{c_1 \phi}$ ($b_1=2$).}
    \label{tab12}
\end{table}

\subsection{Model III}
For Models III and IV, we relax the previous assumption and treat the exponent $b_1$ as a free parameter. We will then assess whether allowing $b_1$ to vary leads to a statistically significant improvement in the fit to cosmological data.

\begin{figure}[htb]
    \centering
    \includegraphics[width=0.8\linewidth]{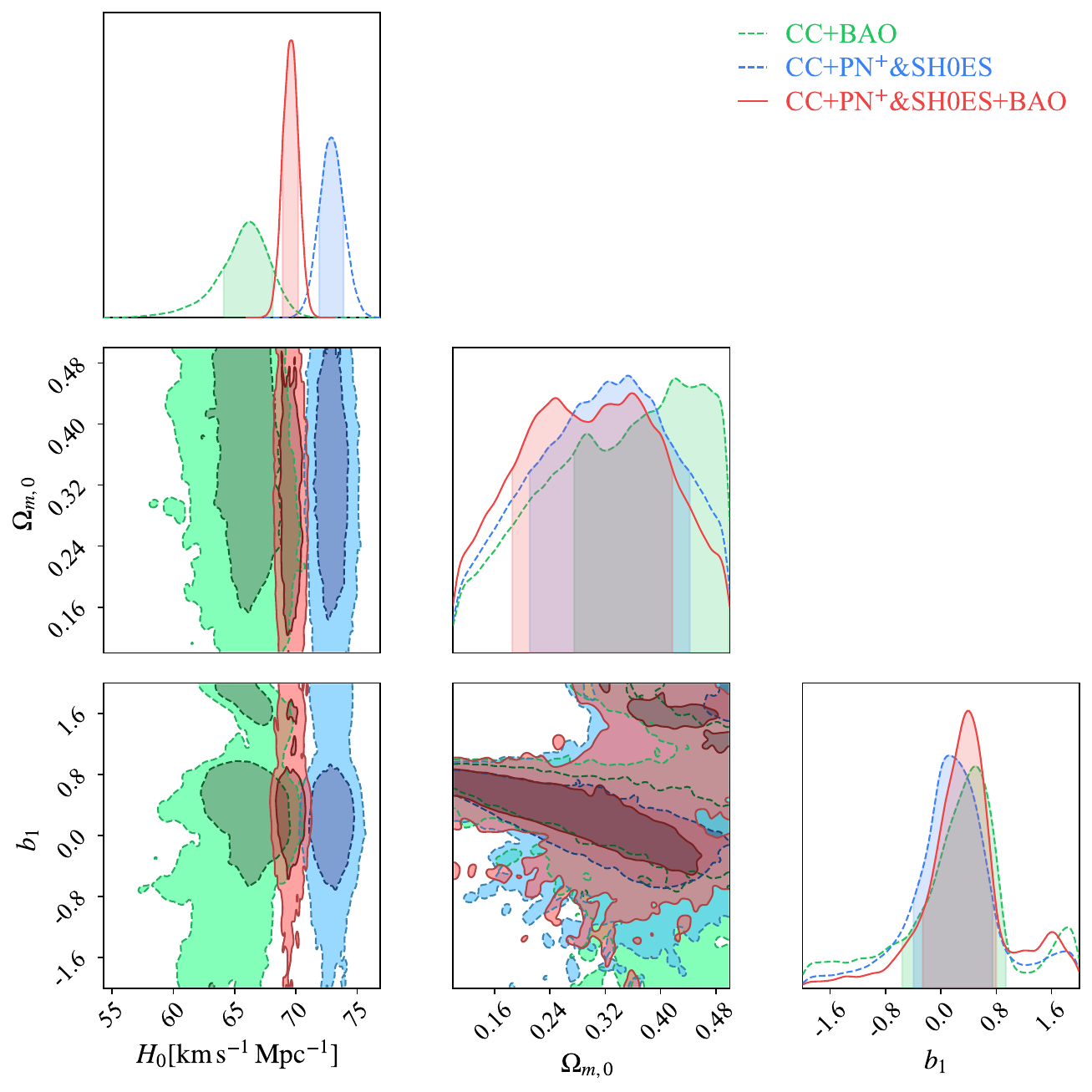}
    \caption{MCMC corner plot for the cubic Galileon teleparallel cosmology with quadratic effective potential $ V(\phi) = c_2 \phi^2$ and $b_1$ left as a free parameter. The contours indicate the 68\% and 95\% marginalized confidence regions, and the diagonal shows the one-dimensional marginalized posteriors.}
    \label{fig:model3}
\end{figure}

The posterior distributions and confidence levels for Model III are presented in Fig.~\ref{fig:model3}. The constraints obtained for this model are summarized in Table~\ref{tab21} for three different data set combinations: CC+BAO, CC+$\text{PN}^+$\&SH0ES, and CC+$\text{PN}^+$\&SH0ES+BAO. When only CC+BAO data are used, the Hubble constant is found to be $H_0 = 66.2^{+1.9}_{-2.2}$ km s$^{-1}$ Mpc$^{-1}$, which is relatively low and comparable to the value favored by early-universe probes. Adding the $\text{PN}^+$\&SH0ES catalog shifts $H_0$ upward significantly, reaching $72.95^{+0.95}_{-1.05}$ km s$^{-1}$ Mpc$^{-1}$ for the CC+$\text{PN}^+$\&SH0ES combination, with a marked reduction in uncertainty. When BAO data are further included, the $H_0$ estimate becomes $69.67^{+0.52}_{-0.76}$ km s$^{-1}$ Mpc$^{-1}$, lying between the previous two values and achieving the highest precision among all combinations. The matter density parameter $\Omega_{m,0}$ shows considerable scatter, with values ranging from $0.42^{+0.078}_{-0.146}$ (CC+BAO) to $0.352^{+0.091}_{-0.142}$ (CC+$\text{PN}^+$\&SH0ES) and $0.357^{+0.059}_{-0.172}$ (CC+$\text{PN}^+$\&SH0ES+BAO). Notably, the free parameter $b_1$ remains poorly constrained across all data sets, with large uncertainties that include zero in all cases, suggesting that the data do not strongly prefer a non-zero value. The nuisance parameter $M$ is well determined when $\text{PN}^+$\&SH0ES data are included, with consistent values around $-19.26$ to $-19.36$. Overall, Model III exhibits similar trends to Models~I and~II, namely that the inclusion of $\text{PN}^+$\&SH0ES raises $H_0$ and improves precision, while the addition of BAO tends to moderate the $H_0$ value and tighten the constraints further.

\begin{table}[htbp]\small
    \centering
    \begin{tabular}{l c c c c c c c}
        \hline\hline
        Data sets 
        & $H_0 \,[\mathrm{km\, s^{-1}\, Mpc^{-1}}]$ & $\Omega_{m,0}$ & $\phi_0$ & $\dot{\phi}_0$ & $b_1$ & $c_2$ & $M$ \\
        \hline
        ${\rm CC}$& $64.4^{+12.1}_{-5.5}$ &$0.38^{+0.11}_{-0.11}$&$0.00^{+0.73}_{-0.75}$&$-$& $-$ & $-$&$-$\\
        ${\rm CC+BAO}$& $66.2^{+1.9}_{-2.2}$ &$0.42^{+0.078}_{-0.146}$&$-0.08^{+0.71}_{-0.72}$&$0.0^{+0.035}_{-0.032}$& $0.5^{+0.44}_{-1.06}$ & $-0.9^{+3.9}_{-6.2}$&$-$\\
        ${\rm CC+PN^+\&SH0ES}$& $72.95^{+0.95}_{-1.05}$ &$0.352^{+0.091}_{-0.142}$&$-0.02^{+0.62}_{-0.65}$&$0.001^{+0.023}_{-0.026}$& $0.10^{+0.65}_{-0.49}$ & $-0.4^{+3.1}_{-6.1}$&$-19.257^{+0.026}_{-0.032}$\\
        ${\rm CC+PN^+\&SH0ES+BAO}$& $69.67^{+0.52}_{-0.76}$&$0.357^{+0.059}_{-0.172}$&$0.00^{+0.80}_{-0.71}$&  $0.000^{+0.023}_{-0.022}$&$0.38^{+0.42}_{-0.65}$&$-0.4^{+3.2}_{-4.9}$&$-19.364^{+0.015}_{-0.020}$\\
        \hline\hline
    \end{tabular}
    \caption{Best-fit values of the cosmological parameters obtained from different observational data sets for the potential $V(\phi)=c_2 \phi^2$.}
    \label{tab21}
\end{table}

\subsection{Model IV}

\begin{figure}[htb]
    \centering
    \includegraphics[width=0.8\linewidth]{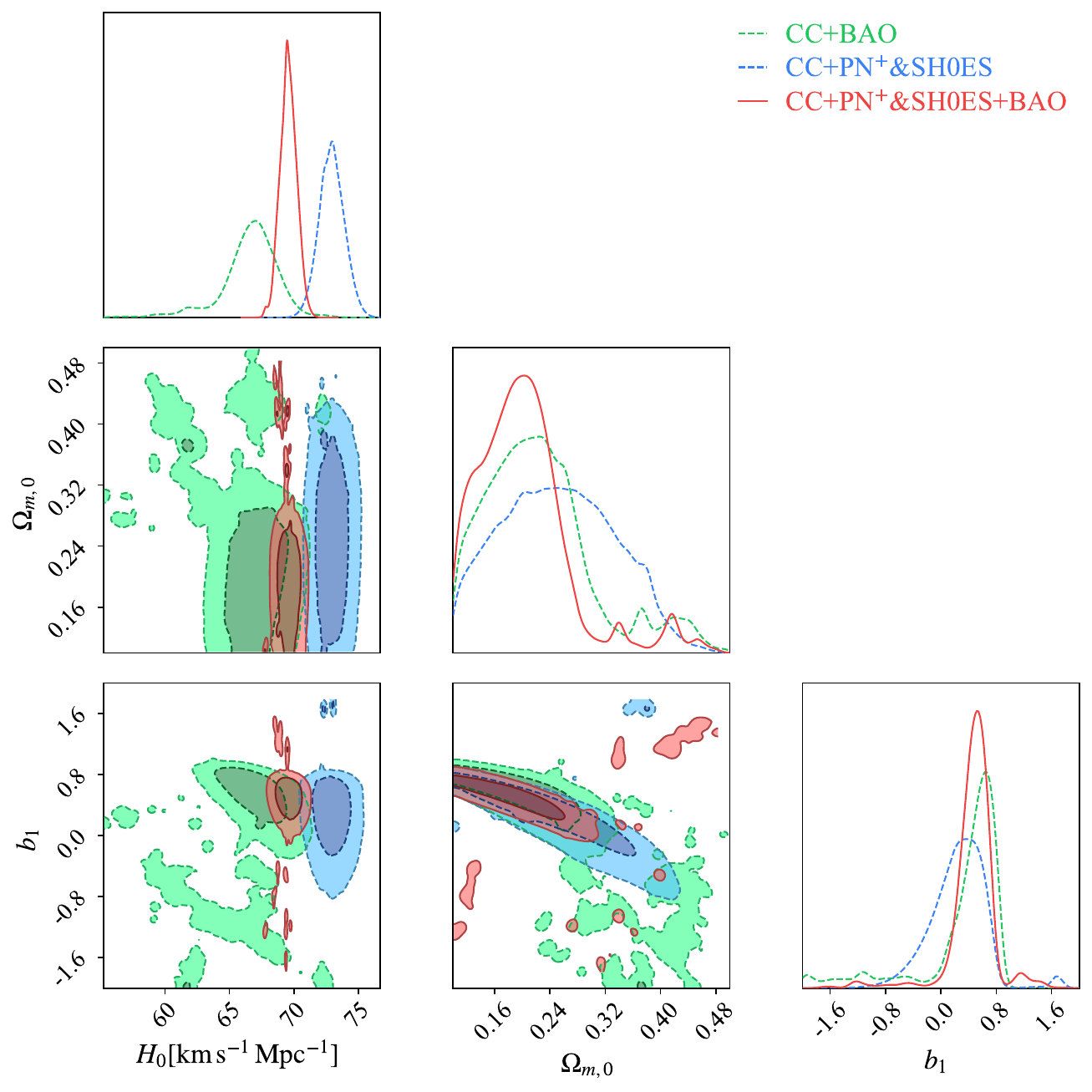}
    \caption{MCMC corner plot for the cubic Galileon teleparallel cosmology with exponential effective potential $ V(\phi) = c_2 e^{c_1 \phi}$ and $b_1$ free. The two-dimensional marginalized contours represent the 68\% and 95\% confidence levels, and the one-dimensional marginalized distributions are plotted on the diagonal.}
    \label{fig:model4}
\end{figure}

The constraints for Model IV are presented in Table~\ref{tab22} and illustrated in Fig.~\ref{fig:model4}. Using only CC+BAO data, the Hubble constant is estimated as $H_0 = 67.1^{+1.7}_{-1.9}$ km s$^{-1}$ Mpc$^{-1}$, with a relatively low central value and moderate uncertainty. When the $\text{PN}^+$\&SH0ES catalog is added (CC+$\text{PN}^+$\&SH0ES), $H_0$ rises significantly to $72.96^{+0.94}_{-1.01}$ km s$^{-1}$ Mpc$^{-1}$, accompanied by a substantial reduction in error bars. Including BAO on top of this combination (CC+$\text{PN}^+$\&SH0ES+BAO) yields an intermediate $H_0$ value of $69.47^{+0.74}_{-0.56}$ km s$^{-1}$ Mpc$^{-1}$, which achieves the highest precision among all three data sets. The matter density parameter $\Omega_{m,0}$ remains consistently low across all combinations, ranging from $0.204^{+0.041}_{-0.080}$ to $0.251^{+0.086}_{-0.098}$, with no clear trend. The free parameter $b_1$ is moderately constrained, showing values around $0.3$–$0.65$ with uncertainties that are relatively smaller than those seen in Model III. The nuisance parameter $M$ is well determined whenever $\text{PN}^+$\&SH0ES data are included, with values of $-19.264^{+0.030}_{-0.026}$ (without BAO) and $-19.366\pm0.017$ (with BAO). Overall, Model IV behaves similarly to the previous models: the inclusion of $\text{PN}^+$\&SH0ES elevates $H_0$ and tightens its uncertainty, while the addition of BAO tempers the $H_0$ estimate and further improves precision. Notably, the exponential potential in Model IV yields a lower $\Omega_{m,0}$ compared to the power-law cases (Models~I and~III), suggesting a distinct background expansion history.

\begin{table}[htbp]\small
    \centering
    \begin{tabular}{l c c c c c c c c}
        \hline\hline
        Data sets 
        & $H_0 \,[\mathrm{km\, s^{-1}\, Mpc^{-1}}]$ & $\Omega_{m,0}$ & $\phi_0$ & $\dot{\phi}_0$ & $b_1$ & $c_1$ & $c_2$ & $M$ \\
        \hline
        ${\rm CC}$& $65.2^{+6.7}_{-5.9}$ &$-$&$-$&$0.000^{+0.035}_{-0.03}$& $0.31^{+0.46}_{-1.51}$ & $-$ & $-$ & $-$\\
        ${\rm CC+BAO}$& $67.1^{+1.7}_{-1.9}$ &$0.227^{+0.049}_{-0.098}$&$-$&$0.001^{+0.018}_{-0.016}$& $0.65^{+0.2}_{-0.36}$ & $-$ & $-$ & $-$\\
        ${\rm CC+PN^+\&SH0ES}$&$72.96^{+0.94}_{-1.01}$&$0.251^{+0.086}_{-0.098}$& $-$ &$0.003^{+0.019}_{-0.027}$&$0.33^{+0.36}_{-0.34}$&$-$&$-$&$-19.264^{+0.030}_{-0.026}$\\
        ${\rm CC+PN^+\&SH0ES+BAO}$&$69.47^{+0.74}_{-0.56}$&$0.204^{+0.041}_{-0.080}$ & $-$ &$(4.7^{+8.3}_{-20.1})\times10^{-3}$&$0.53^{+0.17}_{-0.21}$&$-$&$-$&$-19.366\pm0.017$\\
        \hline\hline
    \end{tabular}
    \caption{Best-fit values of the cosmological parameters obtained from different observational data sets for the potential $V(\phi) = c_2 e^{c_1 \phi}$.}
    \label{tab22}
\end{table}

\section{Comparison of different models}\label{sec:5}

To compare the relative performance of different models, we employ two well-established information criteria: the Akaike Information Criterion (AIC) and the Bayesian Information Criterion (BIC) \citep{1100705,Schwarz:1978tpv}. These criteria balance the goodness of fit against model complexity, penalizing the introduction of additional parameters to avoid overfitting.

They are defined as
\begin{subequations}
\begin{align}
\mathrm{AIC} &= -2 \ln \mathcal{L}_{\mathrm{max}} + 2p, \\\nonumber
\mathrm{BIC} &= -2 \ln \mathcal{L}_{\mathrm{max}} + p \ln N,
\end{align}
\end{subequations}
where $\mathcal{L}_{\mathrm{max}}$ is the maximum likelihood value achieved by the model, $p$ is the number of free parameters, and $N$ is the number of data points. The model with the lowest AIC and BIC values is considered the most parsimonious, i.e., it provides the best fit with the fewest parameters.

For each data set, we identify the $\Lambda$CDM model as the one with the smallest AIC and BIC values, denoted $\mathrm{AIC}_0$ and $\mathrm{BIC}_0$. The other models are then compared to this reference by computing the differences
\begin{equation}
\Delta\mathrm{AIC} = \mathrm{AIC} - \mathrm{AIC}_0, \qquad \Delta\mathrm{BIC} = \mathrm{BIC} - \mathrm{BIC}_0.
\end{equation}
The interpretation of these differences follows standard guidelines \citep{1100705}:

\begin{itemize}
\item $\Delta\mathrm{AIC}$ or $\Delta\mathrm{BIC} \in [0, 3]$: weak evidence against the model being compared (i.e., models are statistically indistinguishable);
\item $\Delta\mathrm{AIC}$ or $\Delta\mathrm{BIC} \in (3, 6]$: mild evidence against the model;
\item $\Delta\mathrm{AIC}$ or $\Delta\mathrm{BIC} > 6$: strong evidence against the model.
\end{itemize}

Equivalently, these differences indicate the level of support for the reference model over the alternative.
\begin{table}[htbp]
\centering
\renewcommand{\arraystretch}{1.2}   % row height
\setlength{\tabcolsep}{10pt}         % column padding

\begin{tabular}{l||ccc|ccc}
\hline
Model & \multicolumn{3}{c|}{${\rm CC+PN^+\&SH0ES}$} & \multicolumn{3}{c}{${\rm CC+PN^+\&SH0ES+BAO}$} \\
\cline{2-7}
& $\chi^2_{\min}$ & $\Delta \mathrm{AIC}$ & $\Delta \mathrm{BIC}$ 
& $\chi^2_{\min}$ & $\Delta \mathrm{AIC}$ & $\Delta \mathrm{BIC}$ \\
\hline
$\Lambda$CDM  & 1531.37 & 0 & 0 & 1568.52 & 0  & 0  \\
Model I    & 1530.44 & 5.07 & 21.44  & 1571.59 & 9.07  & 25.48 \\
Model II    & 1532.25 & 8.88 & 30.71 & 1563.12 & 2.60 & 24.47 \\
Model III     & 1530.63 & 7.26 & 29.09 & 1561.97 & 1.46 & 23.33 \\
Model IV    & 1530.67 & 9.30 & 36.59 & 1562.04 & 3.52 & 30.86 \\
\hline
\end{tabular}
\caption{Comparison of the models for the two data combinations.}
\label{tab:model_comparison}
\end{table}

The model comparison indicates that $\Lambda$CDM remains the statistically preferred scenario overall, particularly according to the Bayesian information criterion. For the ${\rm CC+PN^+\&SH0ES}$ dataset, the cubic Galileon teleparallel models provide only a marginal reduction in $\chi^2_{\min}$ relative to $\Lambda$CDM, which is insufficient to compensate for the larger number of model parameters. Consequently, both AIC and BIC disfavor the extended models in this data combination.
When BAO data are included, the free-$b_1$ quadratic-potential case, Model III, becomes the most competitive among the cubic Galileon scenarios. This model reduces $\chi^2_{\min}$ by approximately $6.55$ relative to $\Lambda$CDM and gives $\Delta{\rm AIC}=1.46$, indicating comparable statistical support according to AIC. However, its large positive value of $\Delta{\rm BIC}=23.33$ shows that the improvement in the goodness of fit is still not sufficient to overcome the penalty associated with the additional parameters. Therefore, Model III may be regarded as phenomenologically viable and competitive at the AIC level, but it is not favored over $\Lambda$CDM by the more conservative BIC criterion.

\section{Conclusion}\label{sec:6}

In this work, we investigated observational constraints on cubic Galileon cosmological models formulated within the teleparallel gravity framework. We considered four representative scenarios distinguished by the choice of scalar-field potential and by the treatment of the teleparallel parameter $b_1$. In particular, we studied quadratic and exponential potentials for both fixed $b_1=2$ and free $b_1$ cases. The models were constrained using different combinations of late-time cosmological data, including cosmic chronometer measurements, the Pantheon+ Type Ia supernova sample with SH0ES calibration, and baryon acoustic oscillation data. Our analysis focused on the background expansion history and on the corresponding constraints on $H_0$, $\Omega_{m,0}$, the scalar-field parameters, and the nuisance parameter $M$.

To facilitate a clear comparison across the different models, data set combinations, and prior choices, we present the constraints on each cosmological parameter side by side in the whisker plot of Fig.~\ref{fig:whisker}. The shaded bands in the plot indicate the adopted prior values, allowing the reader to immediately assess how each prior affects the parameter estimates for the individual models. For the fixed-$b_1$ quadratic-potential case, the inclusion of $\rm Pantheon^+$ and SH0ES data significantly shifts the inferred value of the Hubble constant toward the locally measured range, giving $H_0=72.89^{+1.06}_{-0.93}\,{\rm km\,s^{-1}\,Mpc^{-1}}$ for the ${\rm CC+PN^+\&SH0ES}$ combination. However, once BAO measurements are included, the preferred value decreases to $H_0=68.92^{+0.65}_{-0.70}\,{\rm km\,s^{-1}\,Mpc^{-1}}$, showing that BAO data moderate the high-$H_0$ tendency induced by the SH0ES-calibrated supernova sample. A similar behavior is observed for the fixed-$b_1$ exponential-potential model, for which the full data combination gives $H_0=69.57^{+0.62}_{-0.66}\,{\rm km\,s^{-1}Mpc^{-1}}$.

When $b_1$ is treated as a free parameter, the quadratic-potential model gives $H_0=66.2^{+1.9}_{-2.2}\,{\rm km\,s^{-1}\,Mpc^{-1}}$ for ${\rm CC+BAO}$, while the inclusion of Pantheon+ and SH0ES raises this value to $H_0=72.95^{+0.95}_{-1.05}\,{\rm km\,s^{-1}\,Mpc^{-1}}$. Adding BAO to this combination yields an intermediate and more tightly constrained value, $H_0=69.67^{+0.52}_{-0.76}\,{\rm km\,s^{-1}\,Mpc^{-1}}$. The free-$b_1$ exponential-potential model shows the same qualitative trend, with $H_0=67.1^{+1.7}_{-1.9}\,{\rm km\,s^{-1}\,Mpc^{-1}}$ for ${\rm CC+BAO}$ and $H_0=69.47^{+0.74}_{-0.56}\,{\rm km\,s^{-1}\,Mpc^{-1}}$ for the full data combination. These results indicate that the inferred value of $H_0$ is sensitive to both the data combination and the assumed potential, with SH0ES-calibrated supernova data favoring higher values and BAO measurements pulling the constraints toward intermediate values.

The matter density parameter also shows model-dependent behavior. In the free-$b_1$ quadratic-potential model, the full data combination gives $\Omega_{m,0}=0.357^{+0.059}_{-0.172}$, whereas the free-$b_1$ exponential-potential case prefers a lower value, $\Omega_{m,0}=0.204^{+0.041}_{-0.080}$. This suggests that the exponential potential can lead to a distinct background expansion history compared with the quadratic-potential models. In the free-$b_1$ cases, the parameter $b_1$ is only weakly to moderately constrained. For the quadratic potential, the uncertainties remain large and include values close to zero, indicating that the current late-time data do not strongly require a non-zero teleparallel deviation in this case. For the exponential potential, the constraints on $b_1$ are somewhat tighter, particularly when BAO data are included.

We also compared the models using the Akaike information criterion and Bayesian information criterion. For the ${\rm CC+PN^+\&SH0ES}$ data combination, the cubic Galileon teleparallel models provide only a small reduction in $\chi^2_{\min}$ relative to $\Lambda$CDM, which is not sufficient to compensate for their additional free parameters. Consequently, both AIC and BIC favor the minimal $\Lambda$CDM model for this data set. When BAO data are included, the free-$b_1$ quadratic-potential model becomes the most competitive among the cubic Galileon scenarios. In this case, $\chi^2_{\min}$ is reduced from $1568.52$ for $\Lambda$CDM to $1561.97$, and the model gives $\Delta{\rm AIC}=1.46$, indicating comparable support with $\Lambda$CDM according to the AIC criterion. Nevertheless, the corresponding value $\Delta{\rm BIC}=23.33$ remains large, showing that the improvement in goodness of fit is not sufficient to overcome the stronger BIC penalty associated with the extended parameter space.

Overall, our results show that teleparallel cubic Galileon cosmologies can provide viable background-level descriptions of late-time expansion and can lead to different preferred ranges of the Hubble constant depending on the potential and data combination. Among the models considered here, the free-$b_1$ quadratic-potential scenario is the most competitive once BAO data are included, although $\Lambda$CDM remains statistically favored by the more conservative BIC criterion. Therefore, the present analysis does not establish a decisive preference for cubic Galileon teleparallel cosmology over $\Lambda$CDM, but it identifies parameter regions and model choices that remain phenomenologically viable and worthy of further investigation. A full assessment of theoretical viability, including perturbative stability and structure-growth constraints, is left for future work.

\section*{ACKNOWLEDGEMENT}

The research of A.D. and B.A. was funded by the National Natural Science Foundation of China (NSFC) under Grant No. U2541210. O. Y. gratefully acknowledges support from the Erasmus+ Teaching Mobility programme under Project Code 2025-1-MT01-KA171-HED-000331721 (Mobility ID: STA-03) between the University of Malta and the New Uzbekistan University. O. Y. also thanks the University of Malta for its warm hospitality during the visit.

\appendix

\begin{figure}[htb]
    \centering
    \includegraphics[width=0.6\linewidth]{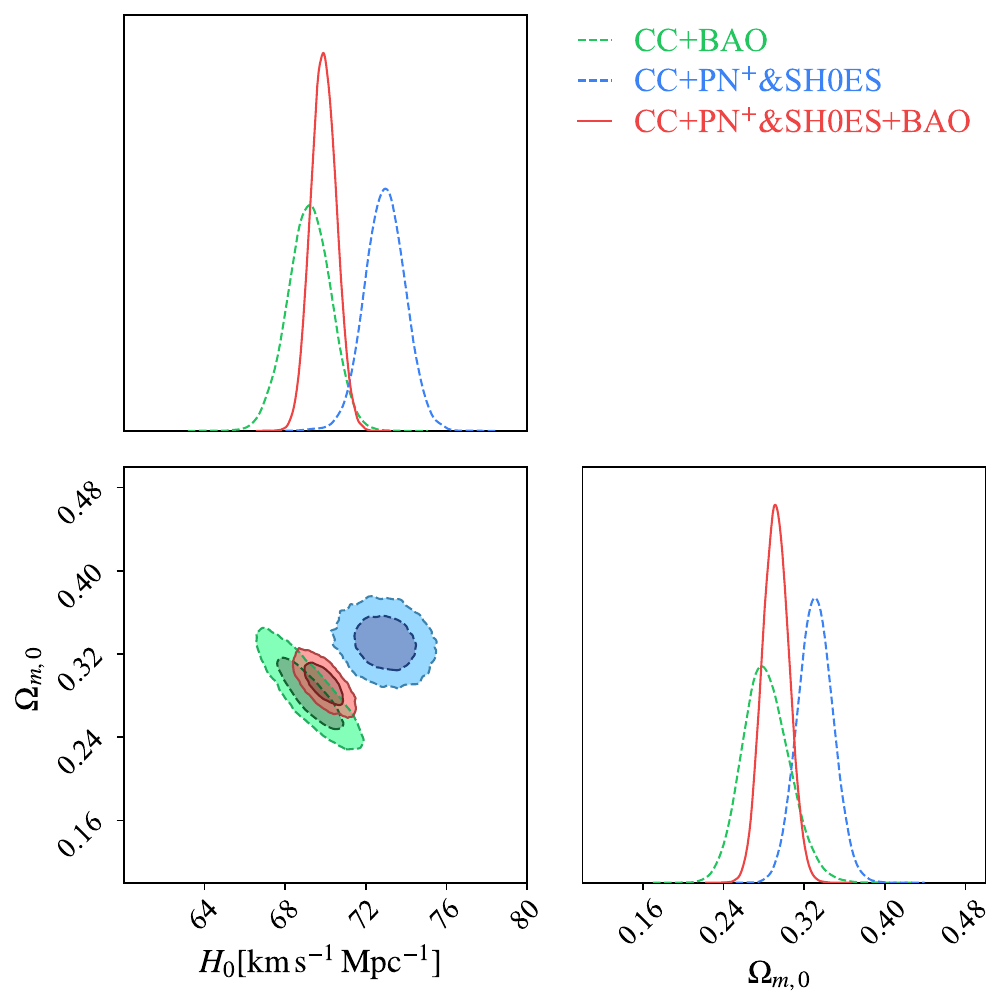}
    \caption{MCMC corner plot for the standard $\Lambda$CDM cosmology, adopted as the reference model for the model comparison analysis. The two-dimensional contours represent the 68\% and 95\% confidence levels of the marginalized posterior distributions, while the diagonal panels display the corresponding one-dimensional marginalized posteriors.}
    \label{2}
\end{figure}

\begin{table}[htbp]\small
    \centering
    \begin{tabular}{l c c c}
        \hline\hline
        Data sets 
        & $H_0 \,[\mathrm{km\, s^{-1}\, Mpc^{-1}}]$ & $\Omega_{m,0}$ & $M$ \\
        \hline
        ${\rm CC}$&$-$ & $-$&$-$\\
        ${\rm CC+BAO}$& $69.3^{+1.0}_{-1.1}$ &$0.278^{+0.026}_{-0.021}$&$-$\\
        ${\rm CC+PN^+\&SH0ES}$& $72.95^{+0.98}_{-0.99}$ &$0.331^{+0.018}_{-0.017}$&$-19.266^{+0.033}_{-0.024}$\\
        ${\rm CC+PN^+\&SH0ES+BAO}$& $69.90^{+0.64}_{-0.63}$&$0.290^{+0.014}_{-0.012}$&$-19.366^{+0.018}_{-0.017}$\\
        \hline\hline
    \end{tabular}
    \caption{Best-fit values of the cosmological parameters obtained from different observational data sets for $\Lambda$CDM.}
    \label{tabLCDM}
\end{table}

\begin{figure}
    \centering
    \includegraphics[width=0.92\linewidth]{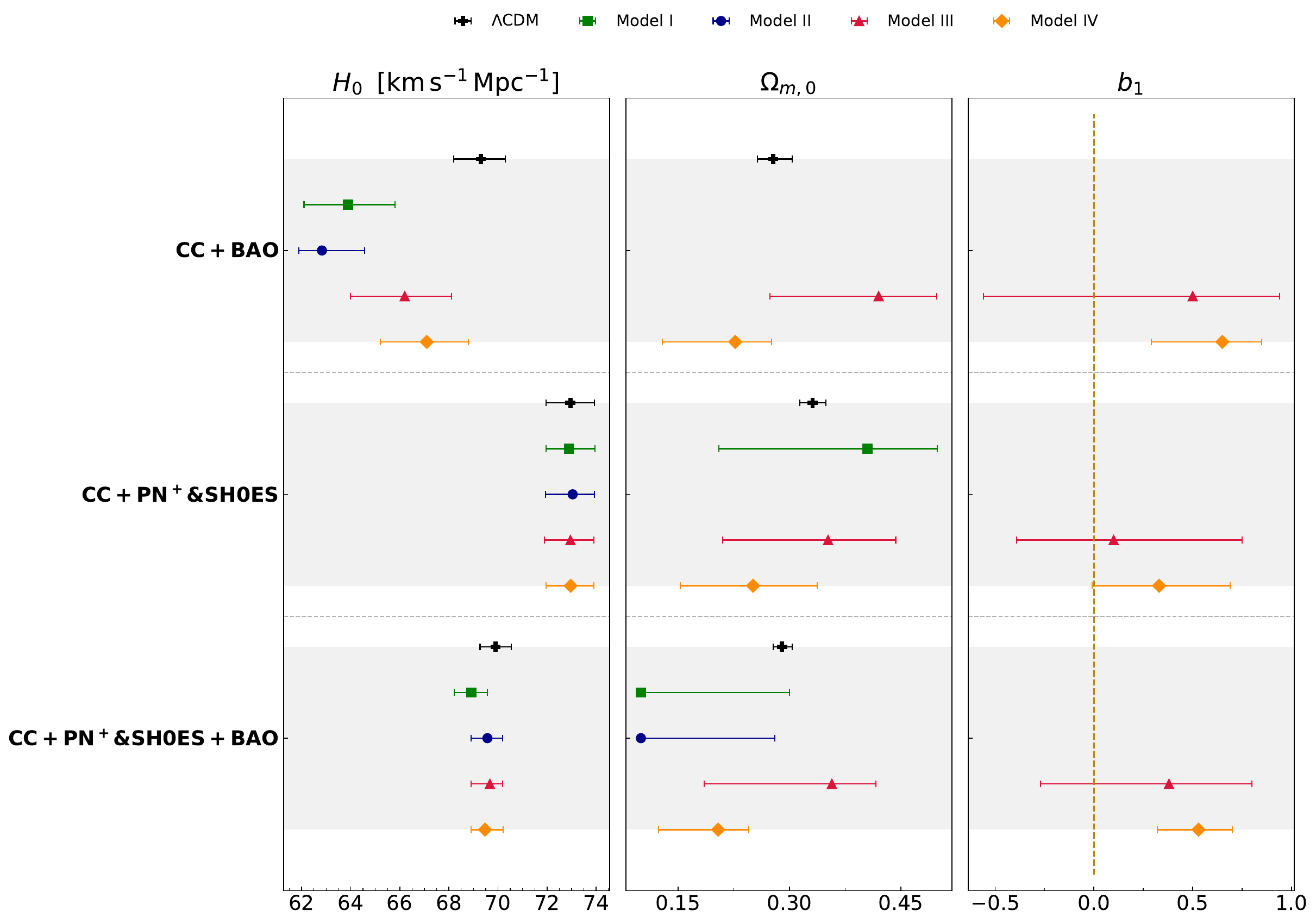}
    \caption{Whisker plot of $H_0$, $\Omega_{m,0}$, and $b_1$, showing the best‑fit values and $1\sigma$ uncertainties for three combinations of data sets (CC+BAO, CC+PN$^+$\&SH0ES, and CC+PN$^+$\&SH0ES+BAO). The models are distinguished by color and best-fit values' marker: $\Lambda$CDM (black, plus), Model I (green, square), Model II (dark blue, circle), Model III (red, triangle), Model IV (orange, diamond).}
    \label{fig:whisker}
\end{figure}

\bibliographystyle{apsrev4-2}
\bibliography{references}

\end{document}